\documentclass[aps,twocolumn,preprintnumbers,showpacs,showkeys,nofootinbib%
]{revtex4}
\usepackage{epsfig}
\usepackage{amssymb,amsmath,amsfonts,amsthm,graphicx,psfrag}
\graphicspath{{./Figures/}}                 
\newcommand{\noi}{\noindent}
\newcommand{\beq}{\begin{equation}}
\newcommand{\eeq}{\end{equation}}
\newcommand{\bea}{\begin{eqnarray}}
\newcommand{\eea}{\end{eqnarray}}
\newcommand{\Fig}[1]{Fig.~\ref{#1}}
\newcommand{\Tab}[1]{Table~\ref{#1}}
\newcommand{\Sec}[1]{Section~\ref{#1}}
\newcommand{\Eq}[1]{Eq.~(\ref{#1})}

\def\ds{\displaystyle}
\def\U{{U_{x,\mu}}}
\def\Ud{{U^\dagger_{x,\mu}}}

\newcommand{\tr}{\operatorname{Tr}}

\newcommand{\bc}{{\it bc~}}
\newcommand{\fc}{{\it fc~}}
\newcommand{\wc}{{\it wc~}}

\newcommand{\aleq}{\mbox{}_{\textstyle \sim}^{\textstyle < }}

\begin{document}

\title{
Infinite volume and continuum limits for gluon propagator in
       $3d \, SU(2)$ lattice gauge theory.
}

\author{V.~G.~Bornyakov}
\affiliation{Institute for High Energy Physics, 142281, Protvino, Russia \\
and Institute of Theoretical and Experimental Physics, 117259 Moscow, Russia}

\author{V.~K.~Mitrjushkin}
\affiliation{Joint Institute for Nuclear Research, 141980 Dubna, Russia \\
and Institute of Theoretical and Experimental Physics, 117259 Moscow, Russia}

\author{R.~N.~Rogalyov}
\affiliation{Institute for High Energy Physics, 142281, Protvino, Russia }

\date{29.04.2013}

\begin{abstract}
We study the Landau gauge gluon propagator $D(p)$ in the 3d $SU(2)$ lattice
gauge theory. We show that in the infinite-volume limit the expectation
values over the Gribov region $\Omega$ are {\it different} (in
the infrared) from that calculated in the fundamental modular region $\Gamma$.
Also we show that this conclusion does {\it not} change when spacing $a$
tends to zero.

\end{abstract}



\maketitle

\section{Introduction}
\label{sec:introduction}

There are various scenarios of confinement based on infrared behavior
of the gauge dependent propagators.  For example, in the Gribov-Zwanziger
(GZ) confinement scenario \cite{Gribov:1977wm, Zwanziger:1991gz} the
Landau-gauge gluon propagator $D(p)$ at infinite volume is expected to
vanish in the infrared (IR) limit $p\to 0$.  At the same time, a refined
Gribov-Zwanziger (RGZ) scenario \cite{Dudal:2007cw, Dudal:2008sp,
Dudal:2008rm} allows a finite nonzero value of $D(0)$.
The nonperturbative lattice calculations are necessary to check the
validity of each scenario as well as to check the results obtained by
analytical methods, e.g., the (truncated) Dyson-Schwinger equations
(DSE) approach.  The DSE scaling solution predicts that the propagator
tends to zero in the zero-momentum limit \cite{vonSmekal:1997is,
Alkofer:2000wg} in accordance with the GZ-scenario.
Another - decoupling -
solution \cite{Cornwall:1981zr, Fischer:2008uz,Aguilar:2008xm,
Boucaud:2008ji} allows a finite nonzero value of $D(0)$ in conformity
with RGZ-scenario.

The $3d$  $SU(2)$ theory can serve as a useful testground to verify
these predictions. It is also of interest for the studies of
the high-temperature limit of the $4d$ theory.  Last years the $3d$
theory has been numerically studied in a number of papers
\cite{Cucchieri:2003di, Cucchieri:2004mf, Cucchieri:2007md,Cucchieri:2007ta,
Maas:2008ri, Cucchieri:2011ig, Bornyakov:2011fn}.
It has been shown that the propagator has a maximum at momenta about
$350\div400$ MeV
and that zero momentum propagator $D(0)$ does not tend to zero in the
infinite-volume limit.

The Gribov copy problem still remains one of the main difficulties
in computation of the gauge-dependent objects (for $3d$ case see, e.g.,
\cite{Bornyakov:2011fn} and references therein).

The manifold consisting of Gribov copies providing local maxima of the
gauge fixing functional $F(U)$ (defined in \Sec{sec:definitions})
and a semi-positive Faddeev-Popov operator is termed the {\it Gribov
region} $~\Omega$, while that of the global maxima is termed the
{\it fundamental modular region} (FMR) $~\Gamma \subset \Omega~$
\cite{SemenovTyanShanskii}.  Our gauge-fixing procedure is aimed to
approach $~\Gamma$.

In paper \cite{Zwanziger:2003cf} it has been claimed  that although
there are Gribov copies inside Gribov region $\Omega$ , they have no
influence on expectation-values in the thermodynamic limit
i.e. for any gauge noninvariant observable $O$
\beq
\langle O \rangle_\Omega = \langle O \rangle_\Gamma \,.
\label{eq:zwanziger}
\eeq

In our recent paper \cite{Bornyakov:2011fn} we attempted to check this
statement. We calculated gluon propagators $D(p)$ on different lattices
(for $p=0$ as well as for $p\ne 0$) and then extrapolated the values of $D$
in the thermodynamic limit. It has been shown that in the thermodynamic
limit $L\to \infty$ the value of propagator  $D(0)$ clearly depends on
the choice of the gauge copy.

The most of our calculations in \cite{Bornyakov:2011fn} has been performed
at $\beta=4.24$ ($a=0.168$ fm).
The main goals of this paper are ({\bf a}) to find confirmation of our
observations made in \cite{Bornyakov:2011fn} employing different (larger)
values of $\beta$
and ({\bf b}) to draw some definite conclusions about the continuum
limit of the theory.

\vspace{1mm}

In \Sec{sec:definitions} we introduce the quantities to be computed
and give some details of our simulations.
In \Sec{sec:results} we present our numerical results.
Conclusions are drawn in \Sec{sec:conclusions}.

\section{Main definitions and details of the simulation}
\label{sec:definitions}

We consider $3d$ cubic lattice $L^3$ with spacing $a$. To generate
Monte Carlo ensembles of thermalized configurations we use the
standard Wilson action

\beq
S  = \beta \sum_{x,\mu >\nu} \left[ 1 -\frac{1}{2} \tr
\Bigl(U_{x\mu}U_{x+\hat \mu a;\nu} U_{x+\hat \nu
a;\mu}^{\dagger}U_{x\nu}^{\dagger} \Bigr)\right]\,,
\label{eq:action}
\eeq

\noindent where $\beta = 4/g_B^2a$,  $\hat \mu$ is a  vector of unit
length along the $\mu$th coordinate axis and $g_B$ denotes
dimensionful bare coupling.  $U_{x\mu} \in SU(2)$ are the link
variables which transform under local gauge transformations $g_x$
as follows:

\beq
U_{x\mu} \stackrel{g}{\mapsto} U_{x\mu}^{g} = g_x^{\dagger}
U_{x\mu} g_{x+\hat \mu a} \,, \qquad g_x \in SU(2) \,.
\label{eq:gaugetrafo}
\eeq

\noi In \Tab{tab:data_sets} we provide the full information about
the field ensembles used in this investigation.
The scale is set in accordance with \cite{Teper:1998te} where
string tension is $\sqrt{\sigma} = 440$~MeV.

\vspace{1mm}

We study the gluon propagator

\bea
D^{bc}_{\mu\nu}(q) &=& {a^3 \over L^3} \sum_{x, y} \exp
\Big(iqx+{ia\over 2}\, q(\hat \mu - \hat\nu)\Big)\\ \nonumber
&& \langle A^b_\mu(x+y+{\hat \mu a\over 2}) A^c_\nu(y+{\hat \nu a\over 2})
\rangle, \nonumber
\eea

\noindent where the vector potentials $A_\mu^a(x)$ are defined as
follows \cite{Mandula:1987rh} :

\beq
A_\mu\Big(x+{\hat \mu a\over 2}\Big) \equiv \sum_{b=1}^3 A^b_\mu
{\sigma^b\over 2} = {i\over g_Ba}\big( \U - \Ud \big),
\eeq

\noindent and the momenta $q_\mu$ take the values $q_\mu =
2\pi n_\mu/aL$, where $n_\mu$ runs over integers in the range
$-L/2 \leq n_\mu < L/2$.
The gluon propagator can be represented in the form

\begin{displaymath}
D^{bc}_{\mu\nu}(q) = \left\{
\begin{array}{l}
\delta^{bc} \delta_{\mu\nu} \bar D(0), \hspace{24mm} p=0;
\\
\delta^{bc} \Big(\ds \delta_{\mu\nu} - {p_\mu p_\nu \over p^2 }\Big)\;
\bar D(p) , \qquad p\neq 0,
\end{array} \right.
\end{displaymath}

\noindent where $\ds ~p_\mu = {2\over a} \sin {q_\mu a\over 2}~$ and
$ ~p^2 = \sum_{\mu=1}^3 p_\mu^2$.
For $p\neq 0$ one arrives at

\beq\label{eq:QuantityUnderStudyDef}
\bar D(p)= {1\over 6} {1 \over ( La)^3}\  \sum_{\mu=1}^{3} \sum_{b=1}^3
\langle \tilde A_\mu^b(q)\, \tilde A_\mu^b(-q) \rangle,
\eeq

\noindent where

\beq
\tilde A_\mu^b(q) = a^3 \sum_{x} A_\mu^b\Big(x+{\hat \mu a\over 2}
\Big)
\exp\Big(\ iq(x+{\hat \mu a\over 2}) \Big),
\eeq

\noindent and the zero-momentum propagator has the form

\beq
\bar D(0)= {1\over 9} {1 \over ( La)^3}\  \sum_{\mu=1}^{3} \sum_{b=1}^3
\langle \tilde A_\mu^b(0)\, \tilde A_\mu^b(0)   \rangle.
\eeq

In what follows we use the gluon propagator $D(p)$ normalized at
$\mu=2.5$ GeV, so that $p^2D(p) =1$ for $p^2=\mu^2$.

We employ the usual choice of the Landau gauge condition on the lattice
\cite{Mandula:1987rh}

\beq
(\partial A)(x) = {1\over
a}\ \sum_{\mu=1}^3 \Big( A_\mu(x+{\hat\mu a\over 2})
  - A_\mu(x-{\hat\mu a\over 2}) \Big)  = 0 \,
\label{eq:diff_gaugecondition}
\eeq

\noindent which is equivalent to finding a local extremum of the gauge
fixing functional

\beq
F_U(g) = ~\frac{1}{3L^3}\sum_{x\mu}~\frac{1}{2}\tr~U^{g}_{x\mu}
\label{eq:gaugefunctional}
\eeq

\noindent with respect to gauge transformations $~g_x$.

\vspace{2mm}

To fix the gauge we choose for every gauge orbit a representative from
$\Gamma~$ \cite{SemenovTyanShanskii}, i.e. the absolute maximum  of the
gauge fixing functional $F(U)$.  This choice is well consistent with a
non-perturbative PJLZ gauge fixing approach \cite{Parrinello:1990pm,
Zwanziger:1990tn} which presumes that a unique representative of the
gauge orbit needs the global extremum of the chosen gauge fixing
functional.
Also in the case of pure gauge $U(1)$ theory in the Coulomb phase some
of the gauge copies produce a photon propagator with a decay behavior
inconsistent with the expected zero mass behavior \cite{Nakamura:1991ww,
Bornyakov:1993yy, Mitrjushkin:1996fw}.
However, the choice of the global extremum permits
to obtain the massless photon propagator.

For practical purposes, it is sufficient to approach the global maximum
close enough so that the systematic errors due to nonideal gauge fixing
(because of, e.g., Gribov copy effects) are  of the same  magnitude as
statistical errors. This strategy has been checked in a number of papers
on $4d$ and $3d$ theory studies for both $SU(2)$ \cite{Bogolubsky:2005wf,
Bogolubsky:2007bw, Bornyakov:2008yx, Bornyakov:2009ug,
Bornyakov:2010nc, Bornyakov:2011fn} and $SU(3)$ \cite{Bornyakov:2011jm,
Aouane:2011fv} gauge groups.

\vspace{3mm}

The gluon propagator in the deep infrared region can be reliably evaluated
only when the effects of Gribov copies are properly taken into account.
The gauge-fixing algorithm which we use was already successfully employed
in the $4d$ theory at both zero \cite{Bornyakov:2008yx,Bornyakov:2009ug} and
nonzero \cite{Bornyakov:2010nc,Bornyakov:2011jm} temperature.
There are three main ingredients of this algorithm: powerful simulated
annealing algorithm, which proved to be efficient in solving various
optimization problems; the flip transformation of gauge fields, which was
used to decrease both the Gribov-copy and finite-volume effects
\cite{Bornyakov:2008yx,Bornyakov:2009ug,Bornyakov:2010nc};
simulation of a large number of gauge copies for each flip sector in order
to further decrease the effects of Gribov copies.


\begin{table}[h]
\begin{center}
\vspace*{0.2cm}
\begin{tabular}{|c|c|c|c|c|} \hline
\multicolumn{5}{|c|}{$\beta=7.09$~~($a=0.094$~fm)}
\\ \hline
 $~L~$ & $~n_{meas}~$ & $n_{copy}$ & $aL$[fm] &  $p_{min}$[GeV]
\\ \hline
  36   & 900   & 160 &   3.38  &  0.365  \\
  48   & 900   & 160 &   4.51  &  0.274  \\
  56   & 900   & 160 &   5.26  &  0.234  \\
  64   & 1200  & 160 &   6.02  &  0.205  \\
  78   & 900   & 280 &   7.33  &  0.168  \\
  92   & 1000  & 280 &   8.65  &  0.143  \\
 \hline\hline
\multicolumn{5}{|c|}{$\beta=10.21$~~($a=0.063$~fm)}
\\ \hline
 $~L~$ & $~n_{meas}~$ & $n_{copy}$ & $aL$[fm] &  $p_{min}$[GeV]
\\ \hline
   36   & 900  &  160 &  2.27  &  0.546   \\
   48   & 900  &  160 &  3.02  &  0.408   \\
   56   & 900  &  160 &  3.53  &  0.350   \\
   64   & 900  &  160 &  4.03  &  0.306   \\
   96   & 700  &  280 &  6.05  &  0.204   \\
   \hline
\end{tabular}
\end{center}
\caption{Values of lattice size, $L$, number of measurements
$n_{meas}$ and number of gauge copies $n_{copy}$ used throughout
this paper.
}
\label{tab:data_sets}
\end{table}


All details of our gauge fixing procedure can be found, e.g., in
\cite{Bornyakov:2011fn}. For readers convenience, we will describe
it shortly here.

Firstly, we extend the gauge group by the transformations (also
referred to as $Z_2$ flips) defined as follows:

\begin{displaymath}
\label{def:nonperiodicGT}
f_\nu(U_{x,\mu}) = \left\{ \begin{array}{l}  -\; U_{x,\mu} \quad
\mbox{if} \quad \ \mu=\nu \quad \mbox{and}\quad
x_{\mu} = a,   \\
\ \ U_{x,\mu} \quad \mbox{otherwise}  \end{array} \right.
\end{displaymath}
\noi which are the generators of the $Z_2^3$ group leaving the action
(\ref{eq:action}) invariant.
Such flips are equivalent to nonperiodic gauge transformations.
A Polyakov loop directed along the transformed links
and averaged over the $2$-dimensional plane changes its sign.
Therefore, the flip operations combine
the $2^3$ distinct gauge orbits (or Polyakov loop sectors) of strictly
periodic gauge transformations into one larger gauge orbit.

We use the simulated annealing (SA), which has been found
computationally more efficient than the use of the standard
overrelaxation (OR) only \cite{Schemel:2006da, Bogolubsky:2007pq,
Bogolubsky:2007bw}.
The SA algorithm generates gauge transformations
$~g(x)~$ by MC iterations with a statistical weight proportional to
$~\exp{(3V~F_U[g]/T)}~$.
The ``temperature'' $~T~$ is an auxiliary
parameter which is gradually decreased in order to maximize the
gauge functional $~F_U[g]~$.  In the beginning, $~T~$ has to be chosen
sufficiently large in order to allow traversing the configuration space
of $~g(x)~$ fields in large steps.
$~T~$ is decreased with equal step size.  The final SA temperature is fixed
such that during the consecutively applied OR algorithm the violation of
the transversality condition

\beq
{g_Ba^2\over 2}\max_{x,c} \big|(\partial A^c)(x)\big| < \, \epsilon
\label{eq:gaugefixstop}
\eeq

\noi decreases in a more or less monotonous manner for the majority of
gauge fixing trials until the condition (\ref{eq:gaugefixstop}) becomes
satisfied with $\epsilon=10^{-7}$.

To finalize the gauge fixing procedure we apply the OR algorithm with
the standard Los Alamos type overrelaxation. In what follows, this method
is labelled FSA  (``Flipped Simulated Annealing'').

We then take the best copy (\bc) out of many gauge fixed copies obtained
for the given gauge field configuration, i.e., a copy with the maximal
value of the lattice gauge fixing functional $F_U$ as a best estimator of
the {\it global} extremum of this functional.

To demonstrate the effect of Gribov copies, we also consider the gauge
obtained by a random choice of a copy within the first Gribov horizon
(labelled as ``\fc''---first copy), i.e., we take the first copy obtained
by the FSA or SA algorithms.  It is instructive also to compare \bc and \fc
propagators with the worst copy (\wc) propagators which correspond to
the choice of the gauge copy with minimal value of the gauge fixing
functional.

\section{Numerical results}
\label{sec:results}

\begin{figure*}[tbh]
\centering
\includegraphics[width=5.9cm,angle=270]{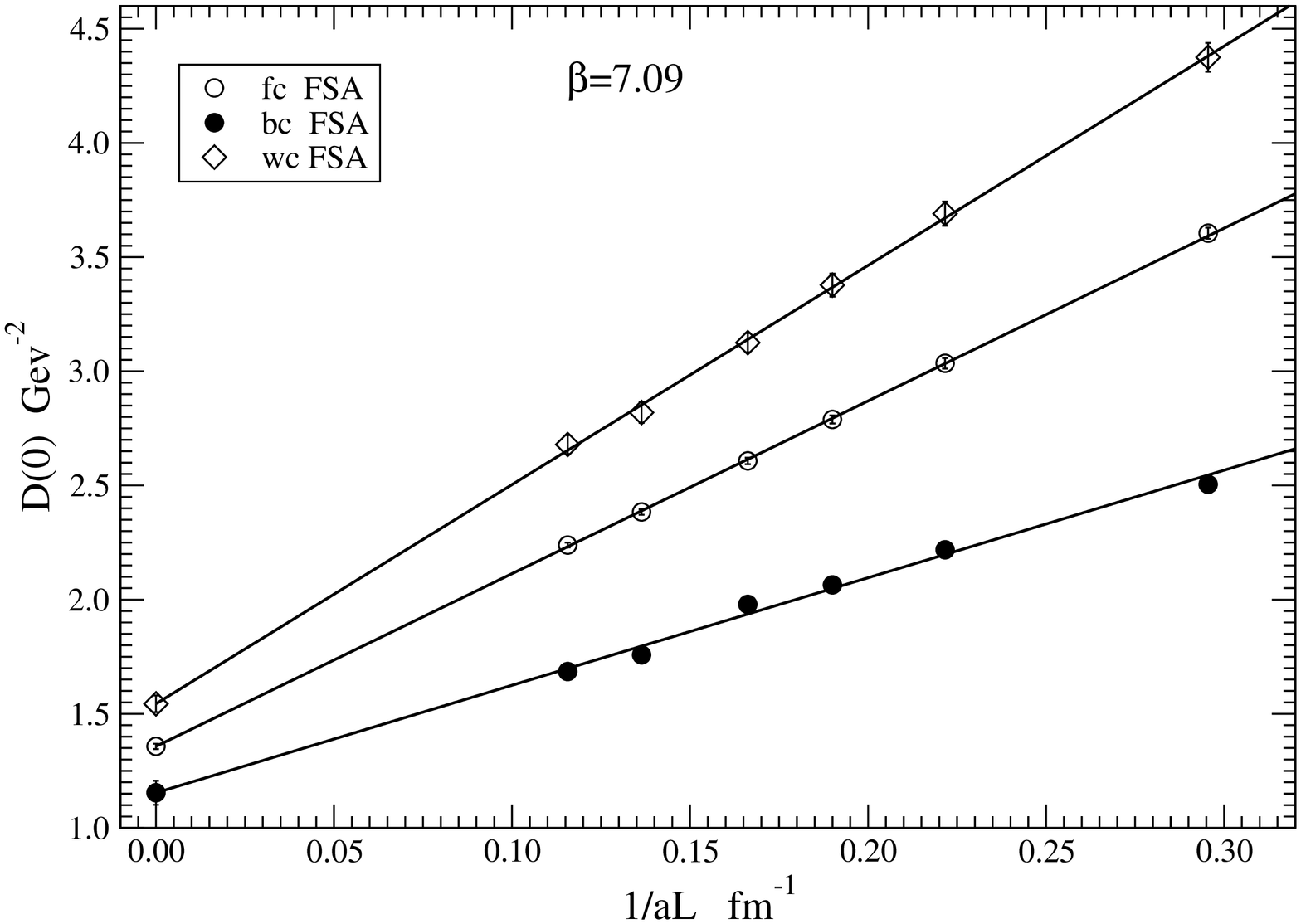}
\includegraphics[width=5.9cm,angle=270]{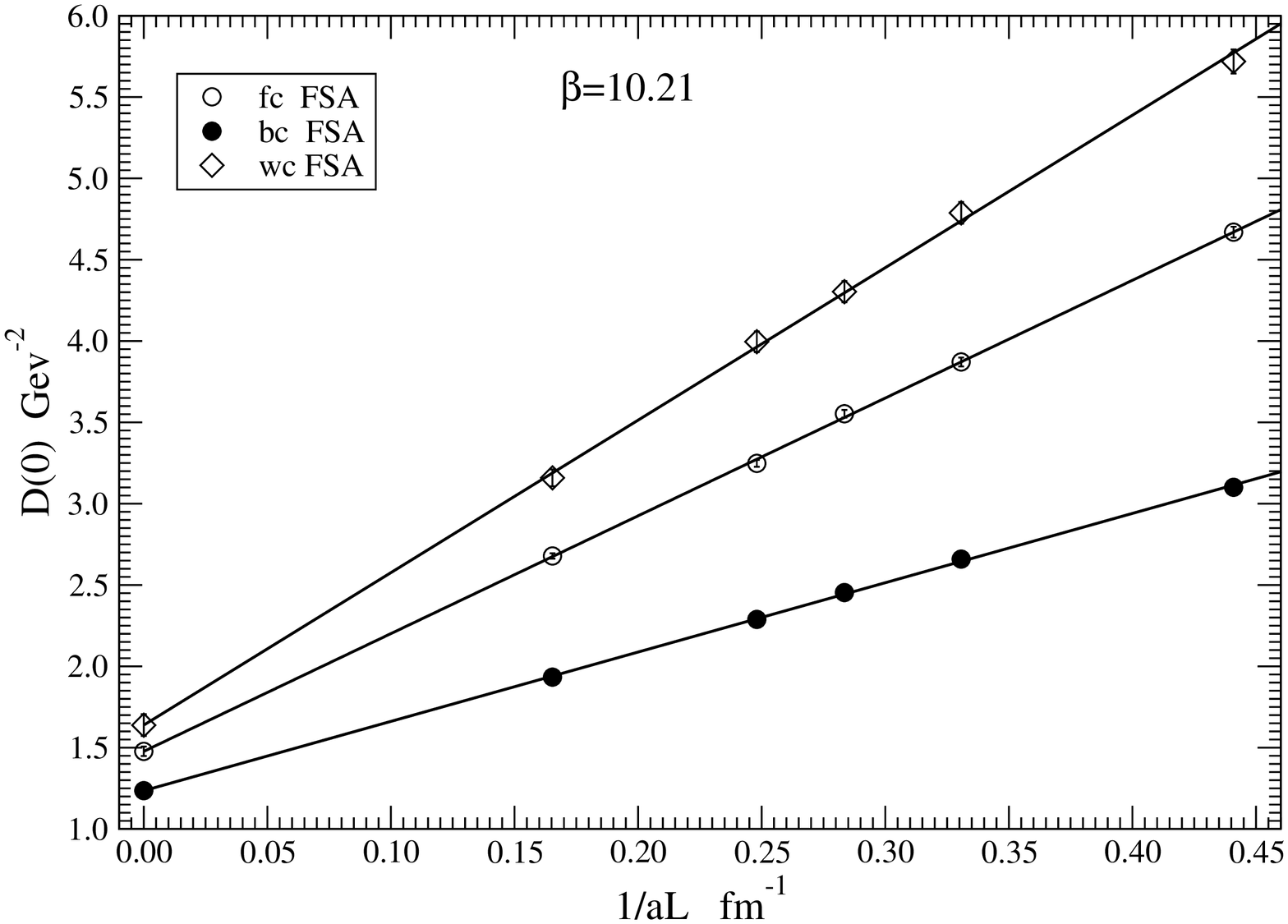}
\caption{$D(0)$ as a function of $1/aL$ for $\beta=7.09$ (Left)
and for $\beta=10.21$ (Right).
}
\label{fig:D0_vs_L_709_1021}
\end{figure*}

\begin{figure*}[tbh]
\centering
\includegraphics[width=5.9cm,angle=270]{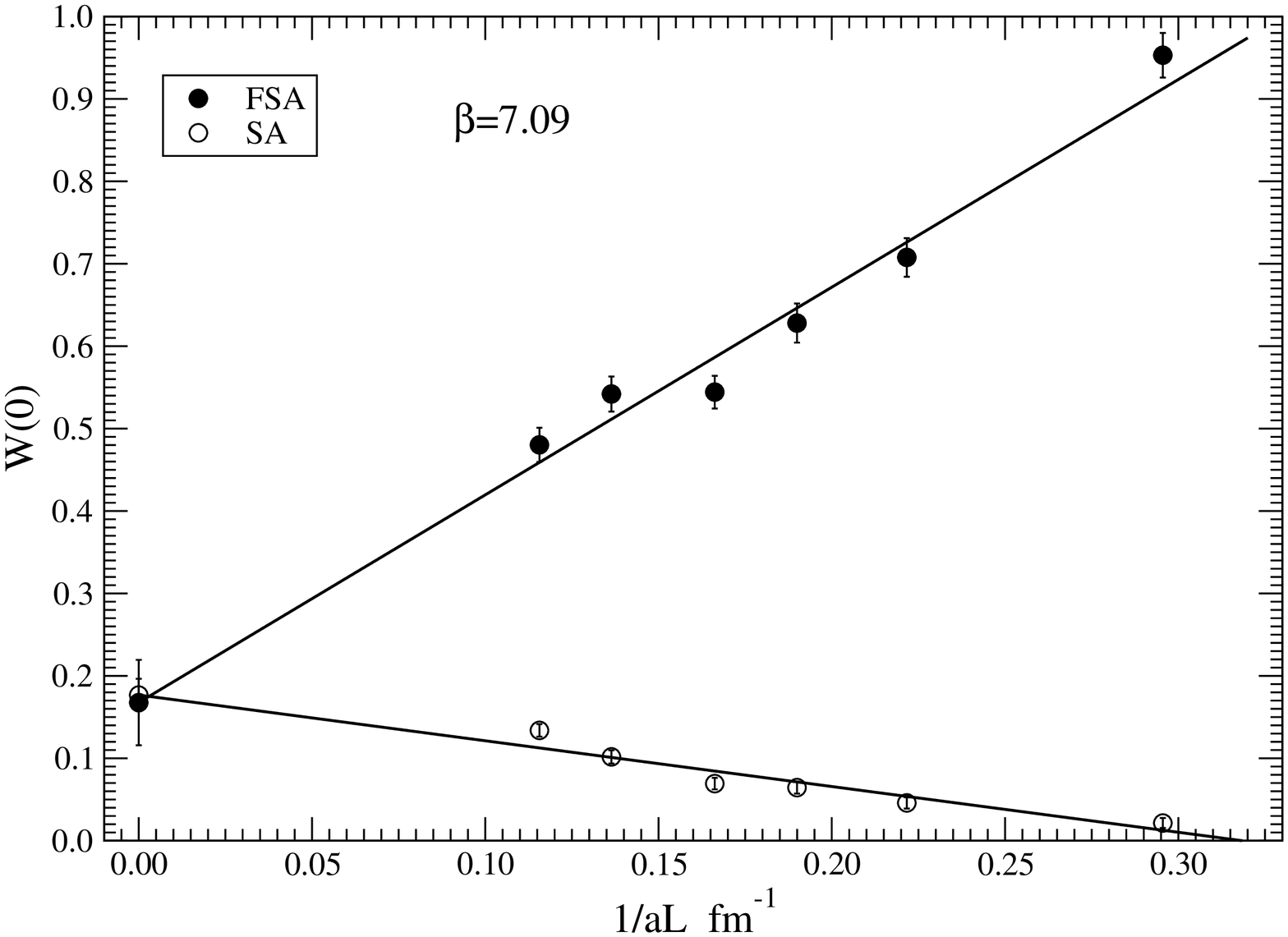}
\includegraphics[width=5.9cm,angle=270]{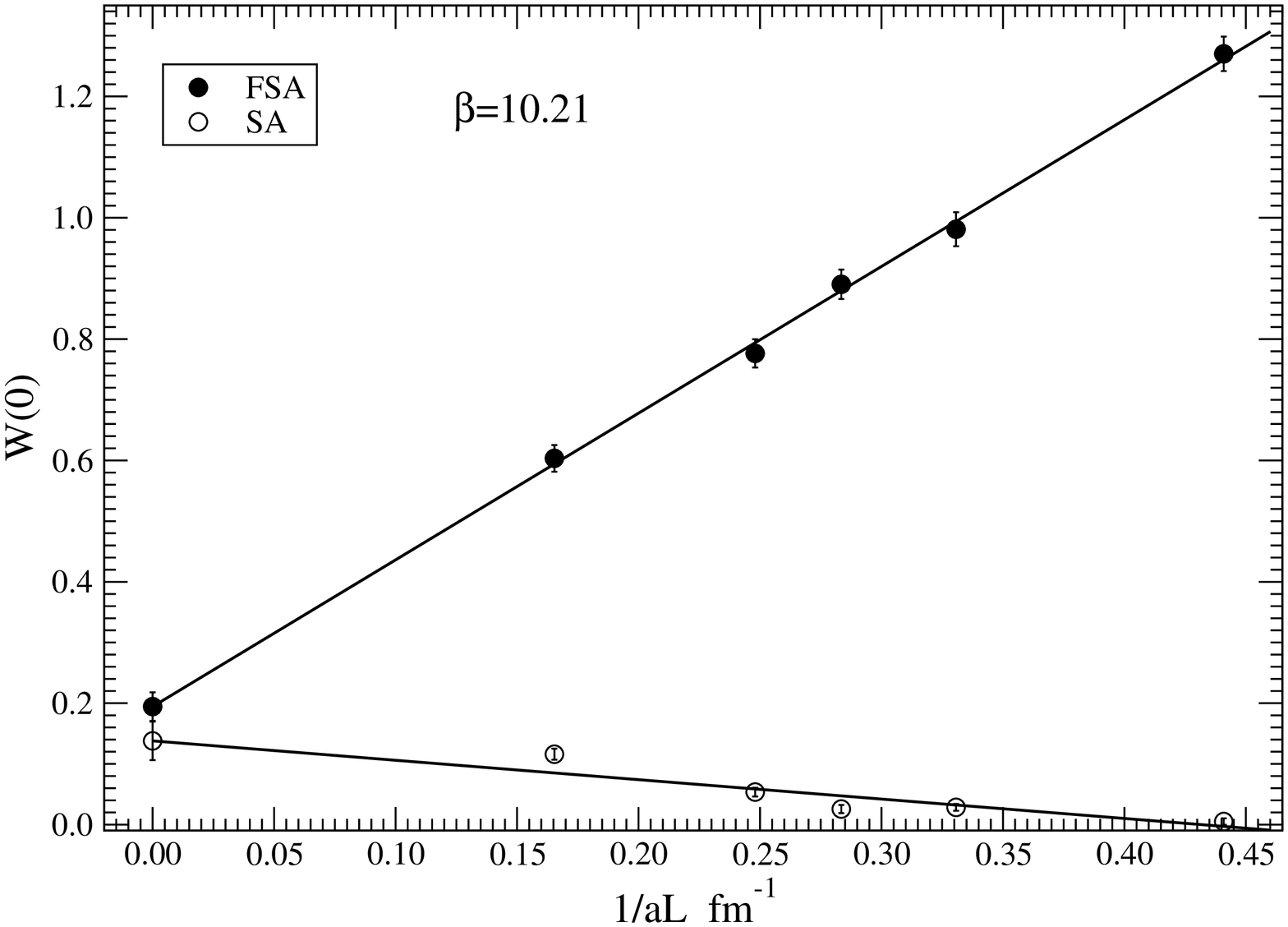}
\caption{$W(0)$ as a function of $1/aL$ for $\beta=7.09$ (Left)  and
$\beta=10.21$ (Right).
}
\label{fig:W0_vs_L_709_1021}
\end{figure*}

\begin{figure}[tb]
\centering
\includegraphics[width=7.1cm,angle=270]{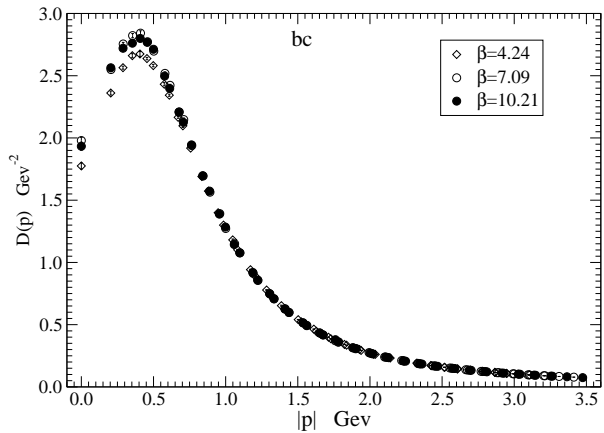}
\caption{\bc $D(p)$ as a function of $|p|$ for $\beta=4.24$,
$\beta=7.09$ and $\beta=10.21$. In all three cases $aL \simeq 6.0$
fm.  Data for $\beta=4.24$ are taken from \cite{Bornyakov:2011fn}.
}
\label{fig:D_vs_p_3beta}
\end{figure}


\begin{figure}[tb]
\centering
\includegraphics[width=7.1cm,angle=270]{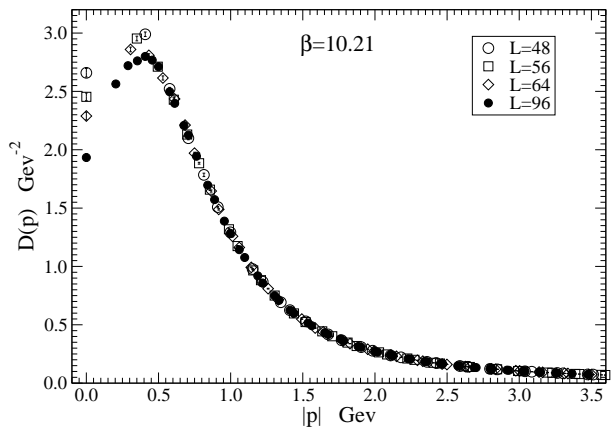}
\caption{Momentum dependence of \bc $D(p)$ for $\beta=10.21$ and
four different volumes.
}
\label{fig:D_vs_p_4Vol}
\end{figure}

\begin{figure*}[tbh]
\centering
\includegraphics[width=5.9cm,angle=270]{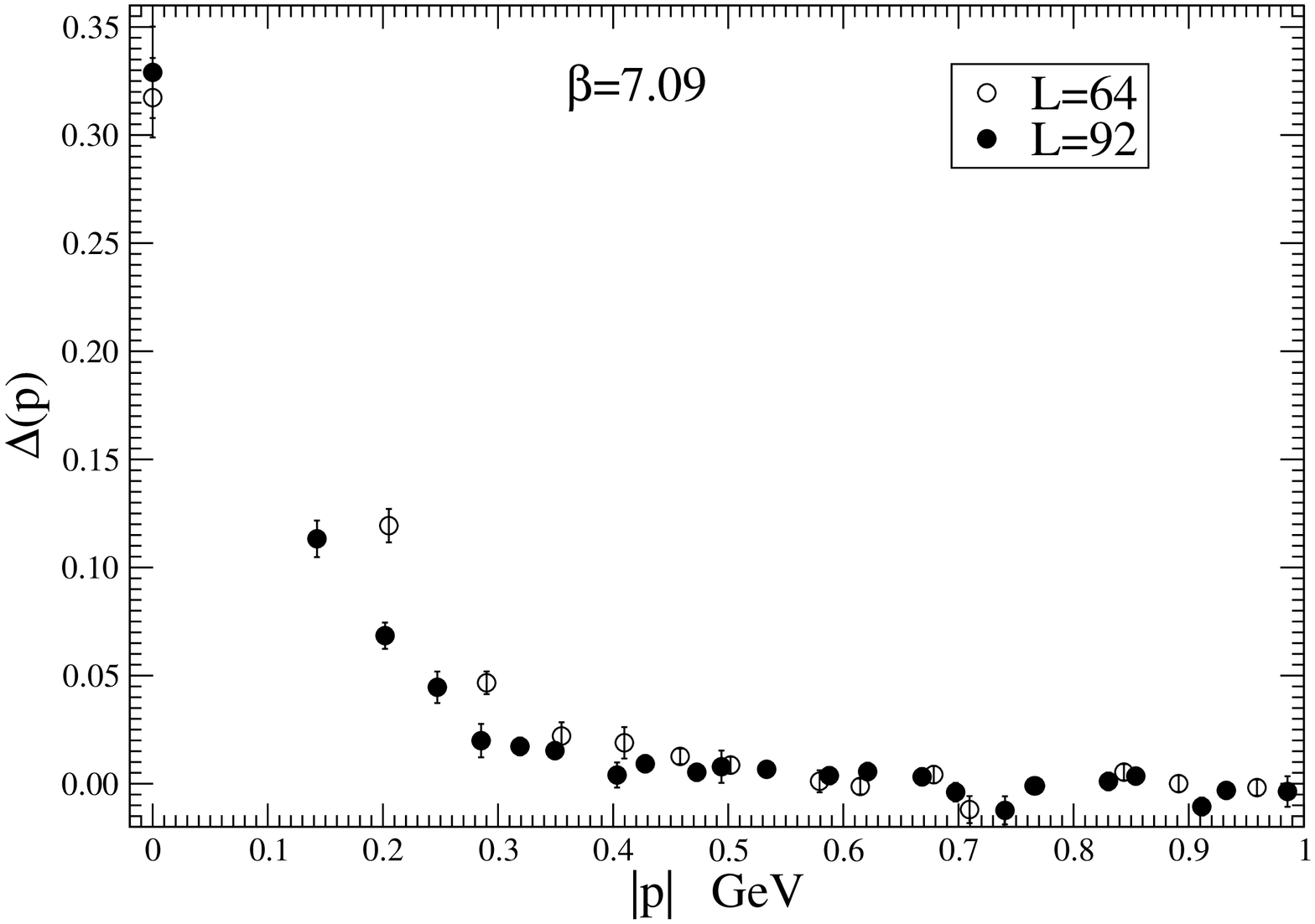}
\includegraphics[width=5.9cm,angle=270]{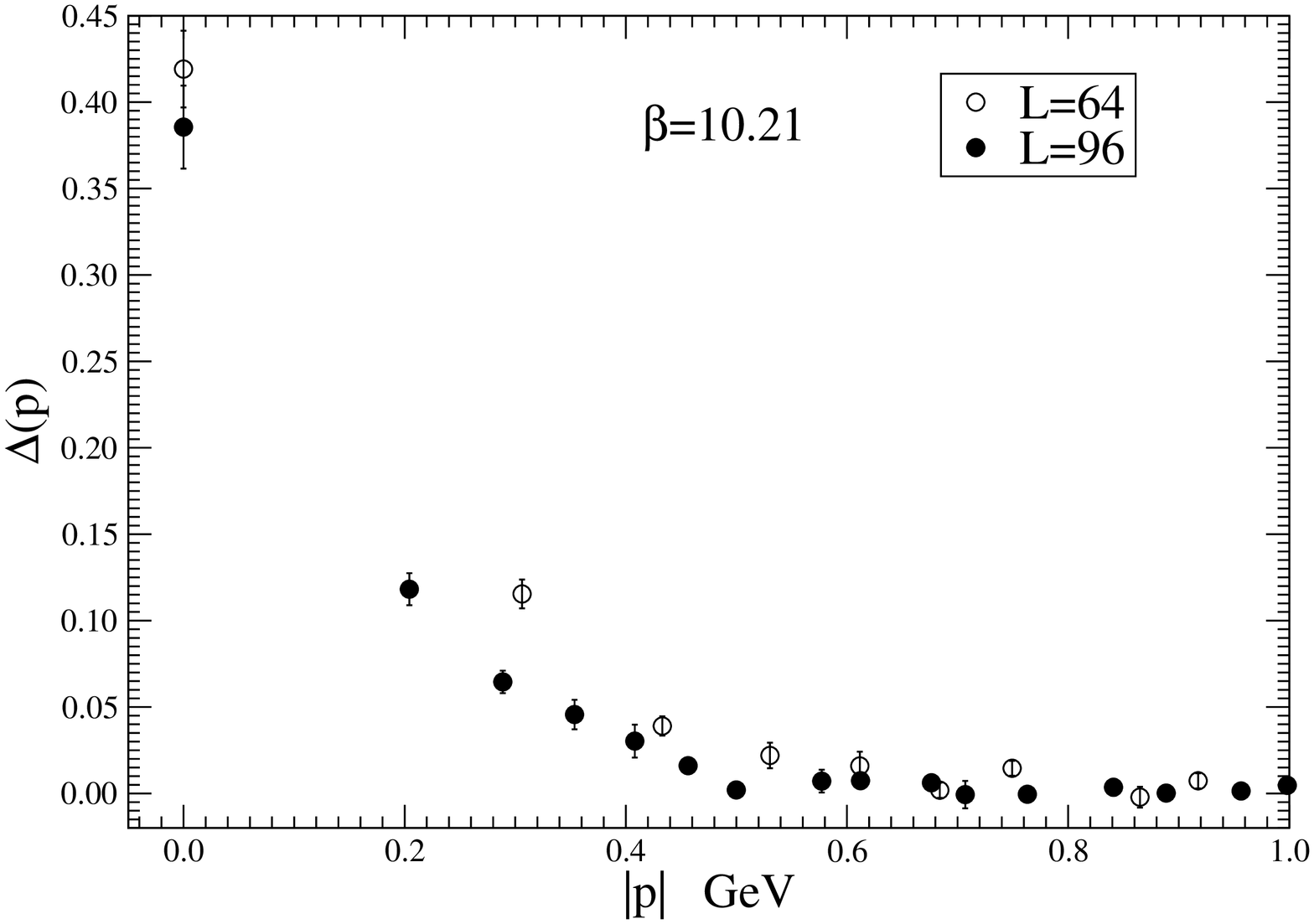}
\caption{$\Delta(p)$ as a function of $p$ for $\beta=7.09$ (Left)
and for $\beta=10.21$ (Right).
}
\label{fig:Delta_vs_p_2beta}
\end{figure*}

To estimate the infinite-volume limit $L\to \infty$ of the zero-momentum
gluon propagator $D(0;L)$ we apply the fit-formula \cite{Cucchieri:2007md}

\beq
D(0;L) =  c_1 + c_2/L~.
\label{eq:fit}
\eeq

In \Fig{fig:D0_vs_L_709_1021} we show our values of $D(0;L)$ calculated
for  $\beta=7.09$ (left panel) and $\beta=10.21$ (right panel) for \bc,
\fc and \wc calculated by FSA method. Straight lines represent fits
according to \Eq{eq:fit}.

In all cases the values of $D(0)$ in the thermodynamic limit {\it
differ} from zero which is in agreement with the statement
made in \cite{Cucchieri:2007md,Bornyakov:2011fn}.

\vspace{1mm}

Moreover, \Fig{fig:D0_vs_L_709_1021} demonstrates another interesting
phenomenon~: the Gribov copy influence remains rather strong even in
the thermodynamic limit.  Indeed, the infinite-volume extrapolation of
$D^{fc}(0)$ {\it differs} from infinite-volume extrapolation of $D^{bc}(0)$
(as well as for $D^{wc}(0)$) which agrees with our observation made in
\cite{Bornyakov:2011fn} for $\beta=4.24$.

Therefore, the expectation values over the Gribov region $\Omega$ are
{\it different} in the infrared from that calculated in the fundamental
modular region $\Gamma$ that disagrees with the statements made in
\cite{Zwanziger:2003cf}.  This is main result of our paper.

As in \cite{Bornyakov:2011fn} we found that
while for finite $L$ $D_{SA}^{bc}(0,L)$ is higher than
$D_{FSA}^{bc}(0,L)$, in the infinite-volume they coincide within errorbars.
This remarkable agreement confirms the
reliability of our estimation of $D(0)$ in the infinite-volume limit.

\vspace{1mm}

For better illustration of our main result we calculated additionally
the averaged difference between \fc and \bc propagators normalized to
$D^{bc}(p;L=\infty)$

\begin{equation}
W(p) = \frac{D^{fc}(p) - D^{bc}(p)}{D^{bc}(p;L=\infty)}
\label{eq:W}
\end{equation}

\noi In \Fig{fig:W0_vs_L_709_1021} we show the dependence of $W(0)$
on the inverse lattice size both for FSA and SA algorithms.
For FSA the value of $W(0)$ decreases for rising size, while for
SA $W(0)$ grows.

For smaller volumes the values of $W(0)$ for SA algorithm, $W_{SA}(0)$,
are close to zero ("Gribov noise") while $W_{FSA}(0)$ is at its
maximum.  The last corresponds to strong effects of flip-sectors (see,
e.g., \cite{Bornyakov:2011fn}).
However, with increasing volume $W_{SA}(0)$ is increasing
indicating the increasing role of the copies within given
flip-sector. In opposite, decreasing of $W_{FSA}(0)$
with increasing volume implies that the role of the flip-sectors
reduces.

Remarkably, in the limit $L\to\infty$ the values
$W_{FSA}(0)$ and $W_{SA}(0)$ coincide (within errorbars) which
confirms the reliability of our fitting procedure.
Results for both algorithms imply {\it nonzero} difference
between \fc and \bc values of the propagators, this difference
being $\sim 15\div 20\%$ in the thermodynamic limit.
Thus effect is very strong.

Note that the asymptotic value $W(0,L=\infty)$ depends weakly
(if any) on the value of the spacing $a$.

\vspace{2mm}

In \Fig{fig:D_vs_p_3beta} we show the momentum dependence of our
\bc gluon propagator $D(p)$ for three different values of $\beta$
(i.e., for three different values of spacing $a$). In all three cases
the physical volumes are approximately equal with $aL \simeq 6.0$ fm.
One can see that the finite-spacing effects are very small if to
compare $\beta=7.09$ and $\beta=10.21$ and are $\sim 1\div 2 \%$
for $|p| \aleq 400$ Mev. For larger values of $|p|$ finite-spacing
effects are even much less. We conclude that (at least) for
$\beta=10.21$ we can speak about the propagator in the continuum limit.

Note that the propagator has a maximum at nonzero value of
momentum  $|p| \sim 400$~MeV. Therefore, the behavior of
$D(p)$ in the deep infrared region is inconsistent with a simple
pole-type dependence.

\vspace{2mm}

In \Fig{fig:D_vs_p_4Vol} we compare the momentum dependence
of the \bc gluon propagator calculated for four different volumes
for $\beta=10.21$.
Apart from $p=0$ case, the finite-volume dependence can be
seen for comparatively small values of momenta. i.e.,
$|p| \aleq  0.5$ GeV. For larger values of momenta the volume
dependence quickly disappears.

\vspace{1mm}

To compare Gribov copy effects for different values of $L$ and
various momenta we define the Gribov copy sensitivity parameter
$\Delta(p) \equiv \Delta(p;L)$ as a normalized difference of the
\fc and \bc gluon propagators

\beq
\Delta(p) = \frac{D^{fc}(p) - D^{bc}(p)}{D^{bc}(p)}~,
\label{delta}
\eeq
\noi where the numerator is the average of the differences
between \fc and \bc propagators calculated for every configuration
and normalized with the \bc (averaged) propagator.

\vspace{2mm}

In \Fig{fig:Delta_vs_p_2beta} we show the momentum dependence of
$\Delta(p)$ for $\beta=7.09$ and $\beta=10.21$ for various volumes.
As one can see, the Gribov copy influence is very strong in deep
infrared and depends weakly on lattice spacing $a$.
For a given value of $L$, the parameter $\Delta(p)$ decreases quickly
with an increase of the momentum.
It is important that for a fixed nonzero physical momentum
$\Delta(p)$ tends to decrease with increasing $L$. Both observations are
in agreement with the observations made earlier in \cite{Bornyakov:2011fn}
and for the four-dimensional $SU(2)$ theory \cite{Bornyakov:2009ug}.

We should emphasize that our results for $p=0$ for all lattice spacings
(see \Fig{fig:D0_vs_L_709_1021} and \Fig{fig:W0_vs_L_709_1021}) as well
as our results for small {\it nonzero} physical momentum presented in
\cite{Bornyakov:2011fn} imply that there are comparatively small
nonzero momenta where Gribov copy effects also survive in the
thermodynamical limit.

The last observation is essential also for the calculation of, e.g.,
screening masses in $4d$ theory at nonzero temperature where the
momentum dependence of the gluon propagator $D(p)$ in the infrared
region is important.

\section{Conclusions}
\label{sec:conclusions}

We investigated numerically the Landau gauge gluon propagator
$D(p)$ in the  $3d$ pure gauge  $SU(2)$ lattice theory.
We have employed lattices with different values of $L$ for $\beta=7.09$
($a=0.094$ fm) and $\beta=10.21$ ($a=0.063$ fm).
This work is the continuation of our previous paper \cite{Bornyakov:2011fn}
where the most of calculations has been done for $\beta=4.24$
($a=0.168$ fm).

The main goal of this work was to confirm our observations made earlier in
\cite{Bornyakov:2011fn} employing larger values of $\beta$
and to draw some definite conclusions about the continuum limit of the theory.

The special attention in this study has been paid to the dependence on the
choice of Gribov copies. To this purpose we have generated up to 280 gauge
copies for every configuration. Our \bc FSA method provides systematically
higher values of the gauge fixing functional as compared to the \fc FSA and
\bc SA methods.
We stress that the choice of the efficient gauge fixing procedure
is of crucial importance in the  study of the gluon propagator in the Landau gauge.

\vspace{1mm}

Our main results are the following.

\begin{itemize}

\item[{\bf 1.}] In the thermodynamic limit $L\to\infty$ the value of $D(0;L)$
differs from zero. This is in agreement with RGZ-scenario and the decoupling
solution of DSE, and confirms the results of numerical computations in
Ref.~\cite{Cucchieri:2007md, Bornyakov:2011fn}.

With decreasing the lattice spacing $a$ the value of $D(0)$ does {\it not} show the
tendency to decrease.

\item[{\bf 2.}] The Gribov copy effects are very strong in the deep
infrared region. Moreover, \fc propagators do {\it not} coincide
with the \bc propagators even in the infinite-volume limit $L\to\infty$
(with difference up to $\sim 15\div 20 \%$).

Therefore, the expectation values over the Gribov region $\Omega$ are
{\it different} in the infrared from that calculated in the fundamental
modular region, i.e.,
\beq
\langle O \rangle_\Omega \ne \langle O \rangle_\Gamma \,.
\label{eq:zwanziger}
\eeq

that does not confirm the statements made
in \cite{Zwanziger:2003cf}.

\item[{\bf 3.}]
The finite-spacing effects appear to be rather small (if to compare
$\beta=7.09$ and $\beta=10.21$).
In the deep infrared the Gribov copy effects for $D(p)$ depend weakly
(if any) on the lattice spacing.
So, we conclude that the difference between averaging over Gribov region
$\Omega$ and fundamental modular region $\Gamma$ persists also in the
continuum limit.

\end{itemize}

\subsection*{Acknowledgments}
This investigation has been partly supported by the Heisenberg-Landau
program of collaboration between the Bogoliubov Laboratory of Theoretical
Physics of the Joint Institute for Nuclear Research Dubna (Russia)
and German institutes, by the grant RFBR 13-02-01387.
VB is supported  by grant RFBR 11-02-01227-a.

\vspace{2mm}


\begin{thebibliography}{36}
\expandafter\ifx\csname natexlab\endcsname\relax\def\natexlab#1{#1}\fi
\expandafter\ifx\csname bibnamefont\endcsname\relax
  \def\bibnamefont#1{#1}\fi
\expandafter\ifx\csname bibfnamefont\endcsname\relax
  \def\bibfnamefont#1{#1}\fi
\expandafter\ifx\csname citenamefont\endcsname\relax
  \def\citenamefont#1{#1}\fi
\expandafter\ifx\csname url\endcsname\relax
  \def\url#1{\texttt{#1}}\fi
\expandafter\ifx\csname urlprefix\endcsname\relax\def\urlprefix{URL }\fi
\providecommand{\bibinfo}[2]{#2}
\providecommand{\eprint}[2][]{\url{#2}}

\bibitem[{\citenamefont{Gribov}(1978)}]{Gribov:1977wm}
\bibinfo{author}{\bibfnamefont{V.~N.} \bibnamefont{Gribov}},
  \bibinfo{journal}{Nucl. Phys.} \textbf{\bibinfo{volume}{B139}},
  \bibinfo{pages}{1} (\bibinfo{year}{1978}).

\bibitem[{\citenamefont{Zwanziger}(1991)}]{Zwanziger:1991gz}
\bibinfo{author}{\bibfnamefont{D.}~\bibnamefont{Zwanziger}},
  \bibinfo{journal}{Nucl. Phys.} \textbf{\bibinfo{volume}{B364}},
  \bibinfo{pages}{127} (\bibinfo{year}{1991}).

\bibitem[{\citenamefont{Dudal et~al.}(2008{\natexlab{a}})\citenamefont{Dudal,
  Sorella, Vandersickel, and Verschelde}}]{Dudal:2007cw}
\bibinfo{author}{\bibfnamefont{D.}~\bibnamefont{Dudal}},
  \bibinfo{author}{\bibfnamefont{S.~P.} \bibnamefont{Sorella}},
  \bibinfo{author}{\bibfnamefont{N.}~\bibnamefont{Vandersickel}},
  \bibnamefont{and}
  \bibinfo{author}{\bibfnamefont{H.}~\bibnamefont{Verschelde}},
  \bibinfo{journal}{Phys. Rev.} \textbf{\bibinfo{volume}{D77}},
  \bibinfo{pages}{071501} (\bibinfo{year}{2008}{\natexlab{a}}),
  \eprint{0711.4496}.

\bibitem[{\citenamefont{Dudal et~al.}(2008{\natexlab{b}})\citenamefont{Dudal,
  Gracey, Sorella, Vandersickel, and Verschelde}}]{Dudal:2008sp}
\bibinfo{author}{\bibfnamefont{D.}~\bibnamefont{Dudal}},
  \bibinfo{author}{\bibfnamefont{J.~A.} \bibnamefont{Gracey}},
  \bibinfo{author}{\bibfnamefont{S.~P.} \bibnamefont{Sorella}},
  \bibinfo{author}{\bibfnamefont{N.}~\bibnamefont{Vandersickel}},
  \bibnamefont{and}
  \bibinfo{author}{\bibfnamefont{H.}~\bibnamefont{Verschelde}},
  \bibinfo{journal}{Phys. Rev.} \textbf{\bibinfo{volume}{D78}},
  \bibinfo{pages}{065047} (\bibinfo{year}{2008}{\natexlab{b}}),
  \eprint{0806.4348}.

\bibitem[{\citenamefont{Dudal et~al.}(2008{\natexlab{c}})\citenamefont{Dudal,
  Gracey, Sorella, Vandersickel, and Verschelde}}]{Dudal:2008rm}
\bibinfo{author}{\bibfnamefont{D.}~\bibnamefont{Dudal}},
  \bibinfo{author}{\bibfnamefont{J.~A.} \bibnamefont{Gracey}},
  \bibinfo{author}{\bibfnamefont{S.~P.} \bibnamefont{Sorella}},
  \bibinfo{author}{\bibfnamefont{N.}~\bibnamefont{Vandersickel}},
  \bibnamefont{and}
  \bibinfo{author}{\bibfnamefont{H.}~\bibnamefont{Verschelde}},
  \bibinfo{journal}{Phys. Rev.} \textbf{\bibinfo{volume}{D78}},
  \bibinfo{pages}{125012} (\bibinfo{year}{2008}{\natexlab{c}}),
  \eprint{0808.0893}.

\bibitem[{\citenamefont{von Smekal et~al.}(1997)\citenamefont{von Smekal,
  Alkofer, and Hauck}}]{vonSmekal:1997is}
\bibinfo{author}{\bibfnamefont{L.}~\bibnamefont{von Smekal}},
  \bibinfo{author}{\bibfnamefont{R.}~\bibnamefont{Alkofer}}, \bibnamefont{and}
  \bibinfo{author}{\bibfnamefont{A.}~\bibnamefont{Hauck}},
  \bibinfo{journal}{Phys. Rev. Lett.} \textbf{\bibinfo{volume}{79}},
  \bibinfo{pages}{3591} (\bibinfo{year}{1997}), \eprint{hep-ph/9705242}.

\bibitem[{\citenamefont{Alkofer and von Smekal}(2001)}]{Alkofer:2000wg}
\bibinfo{author}{\bibfnamefont{R.}~\bibnamefont{Alkofer}} \bibnamefont{and}
  \bibinfo{author}{\bibfnamefont{L.}~\bibnamefont{von Smekal}},
  \bibinfo{journal}{Phys. Rept.} \textbf{\bibinfo{volume}{353}},
  \bibinfo{pages}{281} (\bibinfo{year}{2001}), \eprint{hep-ph/0007355}.

\bibitem[{\citenamefont{Cornwall}(1982)}]{Cornwall:1981zr}
\bibinfo{author}{\bibfnamefont{J.~M.} \bibnamefont{Cornwall}},
  \bibinfo{journal}{Phys.Rev.} \textbf{\bibinfo{volume}{D26}},
  \bibinfo{pages}{1453} (\bibinfo{year}{1982}).

\bibitem[{\citenamefont{Fischer et~al.}(2009)\citenamefont{Fischer, Maas, and
  Pawlowski}}]{Fischer:2008uz}
\bibinfo{author}{\bibfnamefont{C.~S.} \bibnamefont{Fischer}},
  \bibinfo{author}{\bibfnamefont{A.}~\bibnamefont{Maas}}, \bibnamefont{and}
  \bibinfo{author}{\bibfnamefont{J.~M.} \bibnamefont{Pawlowski}},
  \bibinfo{journal}{Annals Phys.} \textbf{\bibinfo{volume}{324}},
  \bibinfo{pages}{2408} (\bibinfo{year}{2009}), \eprint{0810.1987}.

\bibitem[{\citenamefont{Aguilar et~al.}(2008)\citenamefont{Aguilar, Binosi, and
  Papavassiliou}}]{Aguilar:2008xm}
\bibinfo{author}{\bibfnamefont{A.~C.} \bibnamefont{Aguilar}},
  \bibinfo{author}{\bibfnamefont{D.}~\bibnamefont{Binosi}}, \bibnamefont{and}
  \bibinfo{author}{\bibfnamefont{J.}~\bibnamefont{Papavassiliou}},
  \bibinfo{journal}{Phys. Rev.} \textbf{\bibinfo{volume}{D78}},
  \bibinfo{pages}{025010} (\bibinfo{year}{2008}), \eprint{0802.1870}.

\bibitem[{\citenamefont{Boucaud et~al.}(2008)}]{Boucaud:2008ji}
\bibinfo{author}{\bibfnamefont{P.}~\bibnamefont{Boucaud}} \bibnamefont{et~al.},
  \bibinfo{journal}{JHEP} \textbf{\bibinfo{volume}{06}}, \bibinfo{pages}{012}
  (\bibinfo{year}{2008}), \eprint{0801.2721}.

\bibitem[{\citenamefont{Cucchieri et~al.}(2003)\citenamefont{Cucchieri, Mendes,
  and Taurines}}]{Cucchieri:2003di}
\bibinfo{author}{\bibfnamefont{A.}~\bibnamefont{Cucchieri}},
  \bibinfo{author}{\bibfnamefont{T.}~\bibnamefont{Mendes}}, \bibnamefont{and}
  \bibinfo{author}{\bibfnamefont{A.~R.} \bibnamefont{Taurines}},
  \bibinfo{journal}{Phys. Rev.} \textbf{\bibinfo{volume}{D67}},
  \bibinfo{pages}{091502} (\bibinfo{year}{2003}), \eprint{hep-lat/0302022}.

\bibitem[{\citenamefont{Cucchieri et~al.}(2005)\citenamefont{Cucchieri, Mendes,
  and Taurines}}]{Cucchieri:2004mf}
\bibinfo{author}{\bibfnamefont{A.}~\bibnamefont{Cucchieri}},
  \bibinfo{author}{\bibfnamefont{T.}~\bibnamefont{Mendes}}, \bibnamefont{and}
  \bibinfo{author}{\bibfnamefont{A.~R.} \bibnamefont{Taurines}},
  \bibinfo{journal}{Phys. Rev.} \textbf{\bibinfo{volume}{D71}},
  \bibinfo{pages}{051902} (\bibinfo{year}{2005}), \eprint{hep-lat/0406020}.

\bibitem[{\citenamefont{Cucchieri and Mendes}(2007)}]{Cucchieri:2007md}
\bibinfo{author}{\bibfnamefont{A.}~\bibnamefont{Cucchieri}} \bibnamefont{and}
  \bibinfo{author}{\bibfnamefont{T.}~\bibnamefont{Mendes}},
  \bibinfo{journal}{PoS} \textbf{\bibinfo{volume}{LAT2007}},
  \bibinfo{pages}{297} (\bibinfo{year}{2007}), \eprint{0710.0412}.

\bibitem[{\citenamefont{Cucchieri et~al.}(2007)\citenamefont{Cucchieri, Maas,
  and Mendes}}]{Cucchieri:2007ta}
\bibinfo{author}{\bibfnamefont{A.}~\bibnamefont{Cucchieri}},
  \bibinfo{author}{\bibfnamefont{A.}~\bibnamefont{Maas}}, \bibnamefont{and}
  \bibinfo{author}{\bibfnamefont{T.}~\bibnamefont{Mendes}},
  \bibinfo{journal}{Phys. Rev.} \textbf{\bibinfo{volume}{D75}},
  \bibinfo{pages}{076003} (\bibinfo{year}{2007}), \eprint{hep-lat/0702022}.

\bibitem[{\citenamefont{Maas}(2009)}]{Maas:2008ri}
\bibinfo{author}{\bibfnamefont{A.}~\bibnamefont{Maas}}, \bibinfo{journal}{Phys.
  Rev.} \textbf{\bibinfo{volume}{D79}}, \bibinfo{pages}{014505}
  (\bibinfo{year}{2009}), \eprint{0808.3047}.

\bibitem[{\citenamefont{Cucchieri et~al.}(2011)\citenamefont{Cucchieri, Dudal,
  Mendes, and Vandersickel}}]{Cucchieri:2011ig}
\bibinfo{author}{\bibfnamefont{A.}~\bibnamefont{Cucchieri}},
  \bibinfo{author}{\bibfnamefont{D.}~\bibnamefont{Dudal}},
  \bibinfo{author}{\bibfnamefont{T.}~\bibnamefont{Mendes}}, \bibnamefont{and}
  \bibinfo{author}{\bibfnamefont{N.}~\bibnamefont{Vandersickel}}
  (\bibinfo{year}{2011}), \eprint{1111.2327}.

\bibitem[{\citenamefont{Bornyakov et~al.}(2012)\citenamefont{Bornyakov,
  Mitrjushkin, and Rogalyov}}]{Bornyakov:2011fn}
\bibinfo{author}{\bibfnamefont{V.}~\bibnamefont{Bornyakov}},
  \bibinfo{author}{\bibfnamefont{V.}~\bibnamefont{Mitrjushkin}},
  \bibnamefont{and} \bibinfo{author}{\bibfnamefont{R.}~\bibnamefont{Rogalyov}},
  \bibinfo{journal}{Phys.Rev.} \textbf{\bibinfo{volume}{D86}},
  \bibinfo{pages}{114503} (\bibinfo{year}{2012}), \eprint{1112.4975}.

\bibitem[{\citenamefont{Semenov-tyanShanskii and
  Franke}(1982)}]{SemenovTyanShanskii}
\bibinfo{author}{\bibfnamefont{M.~A.} \bibnamefont{Semenov-tyanShanskii}}
  \bibnamefont{and} \bibinfo{author}{\bibfnamefont{V.~A.}
  \bibnamefont{Franke}}, \bibinfo{journal}{{\it Zapiski Nauch. Sem.
  Leningradskogo otd. Matematicheskiogo inst. im. V.A.Steklova}}
  \textbf{\bibinfo{volume}{120}}, \bibinfo{pages}{159} (\bibinfo{year}{1982}),
  \bibinfo{note}{translation: (Plenum, NY, 1986) p.199.}

\bibitem[{\citenamefont{Zwanziger}(2004)}]{Zwanziger:2003cf}
\bibinfo{author}{\bibfnamefont{D.}~\bibnamefont{Zwanziger}},
  \bibinfo{journal}{Phys. Rev.} \textbf{\bibinfo{volume}{D69}},
  \bibinfo{pages}{016002} (\bibinfo{year}{2004}), \eprint{hep-ph/0303028}.

\bibitem[{\citenamefont{Teper}(1999)}]{Teper:1998te}
\bibinfo{author}{\bibfnamefont{M.~J.} \bibnamefont{Teper}},
  \bibinfo{journal}{Phys.Rev.} \textbf{\bibinfo{volume}{D59}},
  \bibinfo{pages}{014512} (\bibinfo{year}{1999}), \eprint{hep-lat/9804008}.

\bibitem[{\citenamefont{Mandula and Ogilvie}(1987)}]{Mandula:1987rh}
\bibinfo{author}{\bibfnamefont{J.~E.} \bibnamefont{Mandula}} \bibnamefont{and}
  \bibinfo{author}{\bibfnamefont{M.}~\bibnamefont{Ogilvie}},
  \bibinfo{journal}{Phys. Lett.} \textbf{\bibinfo{volume}{B185}},
  \bibinfo{pages}{127} (\bibinfo{year}{1987}).

\bibitem[{\citenamefont{Parrinello and Jona-Lasinio}(1990)}]{Parrinello:1990pm}
\bibinfo{author}{\bibfnamefont{C.}~\bibnamefont{Parrinello}} \bibnamefont{and}
  \bibinfo{author}{\bibfnamefont{G.}~\bibnamefont{Jona-Lasinio}},
  \bibinfo{journal}{Phys. Lett.} \textbf{\bibinfo{volume}{B251}},
  \bibinfo{pages}{175} (\bibinfo{year}{1990}).

\bibitem[{\citenamefont{Zwanziger}(1990)}]{Zwanziger:1990tn}
\bibinfo{author}{\bibfnamefont{D.}~\bibnamefont{Zwanziger}},
  \bibinfo{journal}{Nucl. Phys.} \textbf{\bibinfo{volume}{B345}},
  \bibinfo{pages}{461} (\bibinfo{year}{1990}).

\bibitem[{\citenamefont{Nakamura and Plewnia}(1991)}]{Nakamura:1991ww}
\bibinfo{author}{\bibfnamefont{A.}~\bibnamefont{Nakamura}} \bibnamefont{and}
  \bibinfo{author}{\bibfnamefont{M.}~\bibnamefont{Plewnia}},
  \bibinfo{journal}{Phys. Lett.} \textbf{\bibinfo{volume}{B255}},
  \bibinfo{pages}{274} (\bibinfo{year}{1991}).

\bibitem[{\citenamefont{Bornyakov et~al.}(1993)\citenamefont{Bornyakov,
  Mitrjushkin, M{\"u}ller-Preussker, and Pahl}}]{Bornyakov:1993yy}
\bibinfo{author}{\bibfnamefont{V.~G.} \bibnamefont{Bornyakov}},
  \bibinfo{author}{\bibfnamefont{V.~K.} \bibnamefont{Mitrjushkin}},
  \bibinfo{author}{\bibfnamefont{M.}~\bibnamefont{M{\"u}ller-Preussker}},
  \bibnamefont{and} \bibinfo{author}{\bibfnamefont{F.}~\bibnamefont{Pahl}},
  \bibinfo{journal}{Phys. Lett.} \textbf{\bibinfo{volume}{B317}},
  \bibinfo{pages}{596} (\bibinfo{year}{1993}), \eprint{hep-lat/9307010}.

\bibitem[{\citenamefont{Mitrjushkin}(1996)}]{Mitrjushkin:1996fw}
\bibinfo{author}{\bibfnamefont{V.~K.} \bibnamefont{Mitrjushkin}},
  \bibinfo{journal}{Phys. Lett.} \textbf{\bibinfo{volume}{B389}},
  \bibinfo{pages}{713} (\bibinfo{year}{1996}), \eprint{hep-lat/9607069}.

\bibitem[{\citenamefont{Bogolubsky et~al.}(2006)\citenamefont{Bogolubsky,
  Burgio, Mitrjushkin, and M{\"u}ller-Preussker}}]{Bogolubsky:2005wf}
\bibinfo{author}{\bibfnamefont{I.~L.} \bibnamefont{Bogolubsky}},
  \bibinfo{author}{\bibfnamefont{G.}~\bibnamefont{Burgio}},
  \bibinfo{author}{\bibfnamefont{V.~K.} \bibnamefont{Mitrjushkin}},
  \bibnamefont{and}
  \bibinfo{author}{\bibfnamefont{M.}~\bibnamefont{M{\"u}ller-Preussker}},
  \bibinfo{journal}{Phys. Rev.} \textbf{\bibinfo{volume}{D74}},
  \bibinfo{pages}{034503} (\bibinfo{year}{2006}), \eprint{hep-lat/0511056}.

\bibitem[{\citenamefont{Bogolubsky et~al.}(2008)\citenamefont{Bogolubsky,
  Bornyakov, Burgio, Ilgenfritz, Mitrjushkin, and
  M{\"u}ller-Preussker}}]{Bogolubsky:2007bw}
\bibinfo{author}{\bibfnamefont{I.~L.} \bibnamefont{Bogolubsky}},
  \bibinfo{author}{\bibfnamefont{V.~G.} \bibnamefont{Bornyakov}},
  \bibinfo{author}{\bibfnamefont{G.}~\bibnamefont{Burgio}},
  \bibinfo{author}{\bibfnamefont{E.-M.} \bibnamefont{Ilgenfritz}},
  \bibinfo{author}{\bibfnamefont{V.~K.} \bibnamefont{Mitrjushkin}},
  \bibnamefont{and}
  \bibinfo{author}{\bibfnamefont{M.}~\bibnamefont{M{\"u}ller-Preussker}},
  \bibinfo{journal}{Phys. Rev.} \textbf{\bibinfo{volume}{D77}},
  \bibinfo{pages}{014504} (\bibinfo{year}{2008}), \eprint{0707.3611}.

\bibitem[{\citenamefont{Bornyakov et~al.}(2009)\citenamefont{Bornyakov,
  Mitrjushkin, and M{\"u}ller-Preussker}}]{Bornyakov:2008yx}
\bibinfo{author}{\bibfnamefont{V.~G.} \bibnamefont{Bornyakov}},
  \bibinfo{author}{\bibfnamefont{V.~K.} \bibnamefont{Mitrjushkin}},
  \bibnamefont{and}
  \bibinfo{author}{\bibfnamefont{M.}~\bibnamefont{M{\"u}ller-Preussker}},
  \bibinfo{journal}{Phys. Rev.} \textbf{\bibinfo{volume}{D79}},
  \bibinfo{pages}{074504} (\bibinfo{year}{2009}), \eprint{0812.2761}.

\bibitem[{\citenamefont{Bornyakov et~al.}(2010)\citenamefont{Bornyakov,
  Mitrjushkin, and Muller-Preussker}}]{Bornyakov:2009ug}
\bibinfo{author}{\bibfnamefont{V.}~\bibnamefont{Bornyakov}},
  \bibinfo{author}{\bibfnamefont{V.}~\bibnamefont{Mitrjushkin}},
  \bibnamefont{and}
  \bibinfo{author}{\bibfnamefont{M.}~\bibnamefont{Muller-Preussker}},
  \bibinfo{journal}{Phys.Rev.} \textbf{\bibinfo{volume}{D81}},
  \bibinfo{pages}{054503} (\bibinfo{year}{2010}), \eprint{0912.4475}.

\bibitem[{\citenamefont{Bornyakov and Mitrjushkin}(2011)}]{Bornyakov:2010nc}
\bibinfo{author}{\bibfnamefont{V.}~\bibnamefont{Bornyakov}} \bibnamefont{and}
  \bibinfo{author}{\bibfnamefont{V.}~\bibnamefont{Mitrjushkin}},
  \bibinfo{journal}{Phys.Rev.} \textbf{\bibinfo{volume}{D84}},
  \bibinfo{pages}{094503} (\bibinfo{year}{2011}), \eprint{1011.4790}.

\bibitem[{\citenamefont{Bornyakov and Mitrjushkin}(2012)}]{Bornyakov:2011jm}
\bibinfo{author}{\bibfnamefont{V.}~\bibnamefont{Bornyakov}} \bibnamefont{and}
  \bibinfo{author}{\bibfnamefont{V.}~\bibnamefont{Mitrjushkin}},
  \bibinfo{journal}{Int.J.Mod.Phys.} \textbf{\bibinfo{volume}{A27}},
  \bibinfo{pages}{1250050} (\bibinfo{year}{2012}), \eprint{1103.0442}.

\bibitem[{\citenamefont{Aouane et~al.}(2011)\citenamefont{Aouane, Bornyakov,
  Ilgenfritz, Mitrjushkin, Muller-Preussker et~al.}}]{Aouane:2011fv}
\bibinfo{author}{\bibfnamefont{R.}~\bibnamefont{Aouane}},
  \bibinfo{author}{\bibfnamefont{V.}~\bibnamefont{Bornyakov}},
  \bibinfo{author}{\bibfnamefont{E.-M.} \bibnamefont{Ilgenfritz}},
  \bibinfo{author}{\bibfnamefont{V.}~\bibnamefont{Mitrjushkin}},
  \bibinfo{author}{\bibfnamefont{M.}~\bibnamefont{Muller-Preussker}},
  \bibnamefont{et~al.} (\bibinfo{year}{2011}), \eprint{1108.1735}.

\bibitem[{\citenamefont{Schemel}(2006)}]{Schemel:2006da}
\bibinfo{author}{\bibfnamefont{P.}~\bibnamefont{Schemel}},
  \bibinfo{type}{Diploma thesis}, \bibinfo{school}{Humboldt University
  Berlin/Germany} (\bibinfo{year}{2006}).

\bibitem[{\citenamefont{Bogolubsky et~al.}(2007)\citenamefont{Bogolubsky,
  Bornyakov, Burgio, Ilgenfritz, Mitrjushkin, M{\"u}ller-Preussker, and
  Schemel}}]{Bogolubsky:2007pq}
\bibinfo{author}{\bibfnamefont{I.~L.} \bibnamefont{Bogolubsky}},
  \bibinfo{author}{\bibfnamefont{V.~G.} \bibnamefont{Bornyakov}},
  \bibinfo{author}{\bibfnamefont{G.}~\bibnamefont{Burgio}},
  \bibinfo{author}{\bibfnamefont{E.-M.} \bibnamefont{Ilgenfritz}},
  \bibinfo{author}{\bibfnamefont{V.~K.} \bibnamefont{Mitrjushkin}},
  \bibinfo{author}{\bibfnamefont{M.}~\bibnamefont{M{\"u}ller-Preussker}},
  \bibnamefont{and} \bibinfo{author}{\bibfnamefont{P.}~\bibnamefont{Schemel}},
  \bibinfo{journal}{PoS} \textbf{\bibinfo{volume}{LAT2007}},
  \bibinfo{pages}{318} (\bibinfo{year}{2007}), \eprint{0710.3234}.

\end{thebibliography}

\end{document}